%% file: main.tex
\documentclass[conference]{IEEEtran}

\usepackage{graphicx}
\usepackage{textcomp}
\usepackage{xcolor}
\usepackage{color}
\usepackage{graphicx}
\usepackage{url}
\usepackage{cite}
\usepackage{amsthm}
\usepackage{enumerate}
\usepackage{tabularx}
\usepackage{booktabs}
\usepackage{algorithm}
\usepackage{algpseudocode}
\usepackage{multirow}
\usepackage[normalem]{ulem}
\usepackage{tikz}
\usetikzlibrary{shapes,arrows}
\usepackage{amsmath}
\usepackage{amssymb}
\usepackage{colortbl}
\usepackage{xspace}
\usepackage{graphics}
\usepackage{bm}
\usepackage{enumitem}
\usepackage{soul}
\sethlcolor{yellow}
\usepackage{bigstrut}
\usepackage{adjustbox}
\usepackage{comment}
\usepackage{graphicx}

\IEEEoverridecommandlockouts
\makeatletter
\DeclareRobustCommand*\cal{\@fontswitch\relax\mathcal}
\makeatother

\begin{document}
\title{An Open-Source Benchmark Suite of 3D-IC Testcases}

\author{
\IEEEauthorblockN{Rohan Soni, Jooyeon Jeong, Alexander Graening, Anthony Foo, Richard Chen, Puneet Gupta}

\IEEEauthorblockA{
Department of Electrical and Computer Engineering\\
University of California, Los Angeles\\
Los Angeles, CA, USA\\
\{sonir, jooyeon37, agraening, anthonyfoo, chenr4731, puneetg\}@ucla.edu
}
}

\maketitle
\vspace{-5pt}
\input{tex/0_abs.tex}

\input{tex/1_intro.tex}
\input{tex/2_motivation.tex}
\input{tex/3_method.tex}
\input{tex/4_conclusion.tex}

\bibliographystyle{IEEEtran}
\bibliography{main}

\end{document}

%% file: tex/0_abs.tex
\begin{abstract}

The physical design community has benefited from standardized, publicly available benchmark suites, which have enabled reproducible evaluation and driven significant advances in 2D place-and-route algorithms over the past three decades. However, the emergence of 3D heterogeneous integration technologies, including through-silicon vias (TSVs), hybrid bonding, and chiplet-based architectures, has introduced new physical design challenges that are not captured by existing planar benchmarks. Although several 3D-IC design examples have been reported, publicly accessible and scalable benchmark suites that enable reproducible evaluation across different 3D physical design problems remain limited. In this paper, we present an open-source suite of 3D-IC benchmark testcases derived from representative chiplet-based case studies in CATCH \cite{catch2025}, an open-source framework for estimating the cost of heterogeneous integration architectures. The proposed benchmark suite provides reusable virtual chiplet models covering compute, memory, I/O, analog, and substrate components. Each testcase captures essential physical design characteristics of 3D systems, including heterogeneous die integration, inter-die connectivity, and technology-dependent design constraints. By publicly releasing these benchmarks, we aim to establish a common evaluation platform and accelerate community-wide research progress in 3D heterogeneous integration.

\end{abstract}

%% file: tex/1_intro.tex
\section{Introduction}
\label{sec:intro}

Open benchmark suites have played a central role in advancing VLSI physical design by providing common, reproducible targets for algorithm evaluation. The ISPD98 benchmark suite established a widely adopted foundation for placement research, while the ISPD2005 and ISPD2006 placement contests expanded benchmark scale and stimulated advances in congestion-aware and mixed-size placement~\cite{ispd98,ispd2005,ispd2006}. Similarly, the ISPD global-routing benchmarks and ICCAD CAD Contest benchmarks have provided shared evaluation platforms for routing and optimization research~\cite{ispd2008routing,iccad2015}. These experiences demonstrate that representative and openly accessible benchmarks are essential for progress in EDA research.

The history of 2D physical design demonstrates a clear lesson: open and representative benchmarks provide the common evaluation infrastructure required for algorithmic innovation. Without such benchmarks, comparing emerging methodologies becomes difficult. The 3D-IC design community is increasingly facing a similar challenge.

3D heterogeneous integration is rapidly becoming an important technology for continued system scaling, with commercial platforms including AMD 3D V-Cache, NVIDIA GPU--HBM integration, Intel Foveros, TSMC SoIC/CoWoS, and Samsung X-Cube ~\cite{amd_3dv_cache, nvidia2022h100, intel_foveros,tsmc_soic,samsung_xcube} demonstrating different forms of advanced die-to-die and 3D integration. These technologies combine vertically stacked dies, high-density interfaces, interposers, and heterogeneous chiplets, creating design problems that are substantially different from conventional planar physical design.

%% file: tex/2_motivation.tex
\section{Motivation}
\label{sec:motiv}

Despite these rapidly increasing design complexities, publicly available benchmark resources for 3D heterogeneous integration remain limited. Existing examples are often specialized for individual studies, restricted by industrial confidentiality. Consequently, the community lacks a reproducible open benchmark infrastructure that can support fair comparison of emerging 3D physical design methodologies.

Although significant progress has been made toward open-source 3D design methodologies, benchmark resources remain limited. Recent studies have proposed algorithms for monolithic and heterogeneous 3D placement, floorplanning, and routing using representative industrial or academic designs~\cite{pin3d2020,pin3d2024,compact2d2018,liao2024,zhao2025}. The ICCAD CAD Contest has also introduced contest benchmarks for specific 3D placement problems~\cite{phyd3d2025}. More recently, Open3DBench~\cite{open3dbench2025} and RosettaStone~2.0~\cite{rosetastone2026} have established open-source backend implementation flows for 3D physical design based on the OpenROAD ecosystem, enabling reproducible studies of placement, routing, RC extraction, timing, and thermal analysis.

However, these efforts primarily focus on validating physical design methodologies within specific implementation flows. They do not provide a reusable benchmark infrastructure that systematically spans the broad design space encountered in heterogeneous integration, including variations in chiplet composition, die counts, packaging technologies, and cost-driven architectural configurations. As a result, evaluating emerging methodologies across diverse 3D system architectures remains challenging, and fair comparison between competing approaches is often difficult.

We introduce an open-source benchmark suite for heterogeneous 2.5D/3D physical design. The benchmark instances are generated using the open-source CATCH framework~\cite{catch2025}, which provides a convenient representation of heterogeneous chiplet systems and their packaging configurations. The main contributions of this paper are summarized as follows.
 
\begin{itemize}

\item We develop an open-source \textbf{2.5D/3D benchmark infrastructure} comprising (i) a reusable virtual chiplet library for systematic benchmark generation, (ii) an automated \textbf{benchmark conversion framework} that translates CATCH designs into an industry-standard benchmark representation, and (iii) a suite of 20 heterogeneous 2.5D/3D benchmark designs organized into four categories of increasing complexity. The benchmarks are released as standardized 3Dblox descriptions~\cite{ieee3dblox} to enable reproducible evaluation and future benchmark extensions.

\end{itemize}

The remainder of the paper is structured as follows. Section~\ref{sec:method} describes the benchmark methodology. Section~\ref{sec:conclusion} concludes and discusses future use cases.

%% file: tex/3_method.tex
\section{Methodology}
\label{sec:method}

\subsection{Benchmark Generation Framework}

\begin{figure*}[htbp]
    \centering
    \makebox[\textwidth][c]{%
        \hspace{0.5cm}%
        \includegraphics[width=1.15\textwidth]{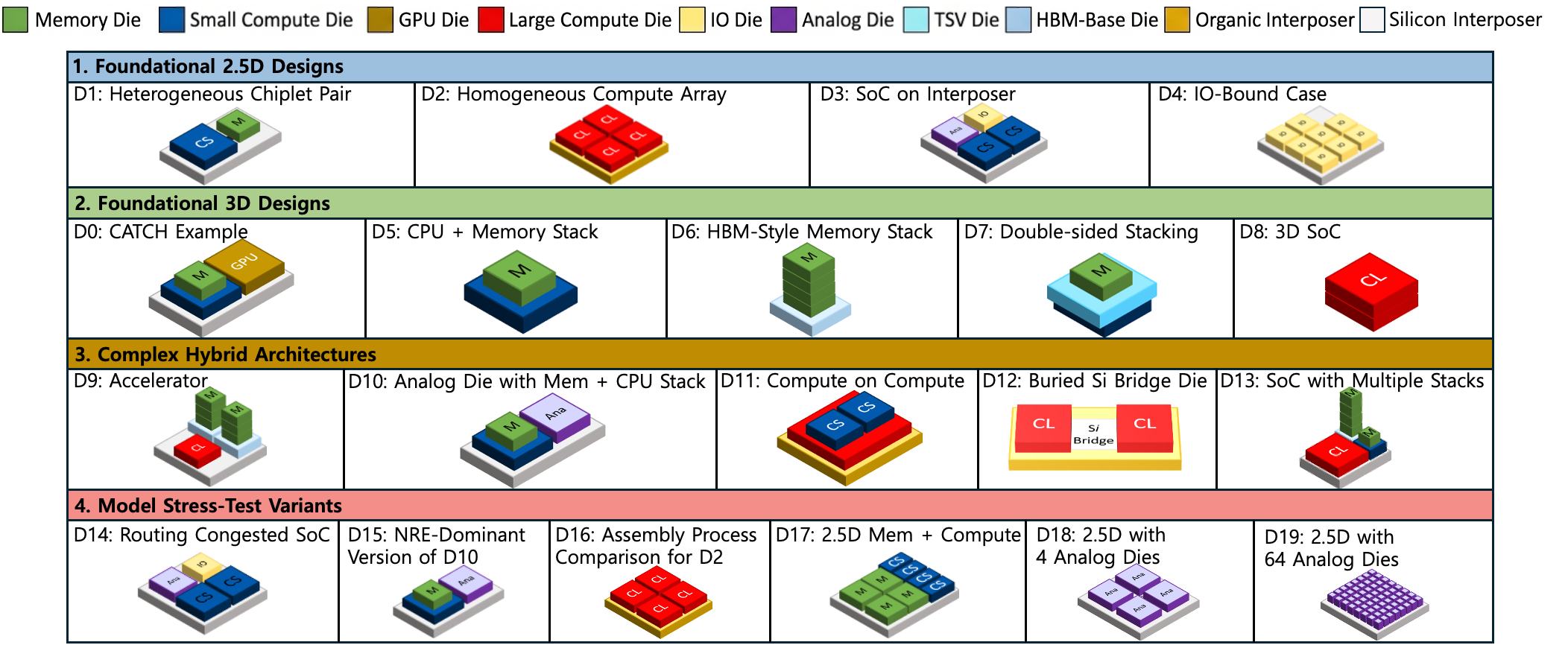}%
    }
    \caption{Visualization of testcases. A legend of chiplets is provided on the left.}
    \label{fig:hbm_area_comp}
\end{figure*}

Our benchmark suite is constructed in a two-phase process: first, we define a reusable chiplet library (Phase~1); second, we compose the library elements into a graded suite of system-in-package designs (Phase~2). All designs are expressed using the CATCH XML schema, which supports arbitrary stack hierarchies, face/back-side connectivity, TSV pass-through signaling, buried dies, and substrate-level routing constraints. This section describes the design intent and structural organization of each benchmark.
 
\subsubsection{(Phase 1) Chiplet Library}\mbox{}\par

\noindent\textbf{Active Dies}
 
We define five virtual active-die templates, each characterized by its CATCH parameters:
 
\begin{itemize}
  \item \textbf{Compute\_Small (CS):}
    A small CPU-core analog with a core area of approximately 20\,mm$^2$ and a logic fraction of 1.0. 
 
  \item \textbf{Compute\_Large (CL):}
    Representing a large GPU or accelerator die, this template has a core area of approximately 250\,mm$^2$ and a logic fraction of 1.0. 
 
  \item \textbf{Memory\_DRAM (MEM):}
    A DRAM die designed for vertical stacking, with a core area of approximately 50\,mm$^2$ and a memory fraction of 1.0. 
 
  \item \textbf{IO\_Controller (IOC):}
    An I/O-pad-dominated chiplet with a core area of approximately 10\,mm$^2$ and a high connection count in the netlist. 
 
  \item \textbf{Analog\_Chiplet (ANA):}
    An analog component template with a core area of approximately 15\,mm$^2$ and an analog fraction of 1.0. It is included to represent the heterogeneous technology nodes commonly found in wireless SoC disaggregation scenarios.
\end{itemize}
 
\noindent\textbf{Passive Substrates}
 
Two substrate templates are defined:
 
\begin{itemize}
  \item \textbf{Substrate\_Organic (SO):}
    A standard, low-cost organic laminate substrate with a large bonding pitch of approximately 110\,$\mu$m. It is used in cost-sensitive designs where I/O density requirements are moderate.
 
  \item \textbf{Substrate\_Silicon (SSi):}
    A high-density silicon interposer with a fine bonding pitch of approximately 45\,$\mu$m, corresponding to advanced CoWoS-style packaging. It enables high-bandwidth die-to-die communication but at a significant area and cost premium.
\end{itemize}
 
\noindent\textbf{Specialty Die}
 
\begin{itemize}
  \item \textbf{TSV\_Interconnect\_Die:}
    A pass-through signaling die defined with connections on both its face and back sides. It is used as the base logic die in HBM-style stacks and to test double-sided stacking configurations.
\end{itemize}
 
\subsubsection{(Phase 2) Benchmark Design Suite}
 
Using the chiplet library, we construct 20 benchmark designs organized into four categories. Table~\ref{tab:designs} summarizes the full suite.
 
\begin{table*}[htbp]
  \centering
  \caption{Summary of the 3D-IC Benchmark Design Suite}
  \label{tab:designs}
  \renewcommand{\arraystretch}{1.2}
  \begin{tabular}{clll}
    \toprule
    \textbf{ID} & \textbf{Name} & \textbf{Chiplet Composition}
                & \textbf{Primary Stress} \\
    \midrule
    \multicolumn{4}{l}{\textit{Category 1 --- Foundational 2.5D Designs}} \\
    D1  & Simple Heterogeneous Pair      & CS + MEM on SSi
        & Basic 2.5D placement \\
    D2  & Homogeneous Compute Array      & 4$\times$CL on SO
        & Multi-chiplet assembly cost scaling \\
    D3  & Full SoC on Interposer         & 2$\times$CS + IOC + ANA on SSi
        & Complex heterogeneous mix \\
    D4  & IO-Bound Stress Test           & 8$\times$IOC on SSi
        & $A_{\mathrm{pads}}$ dominance \\
    \midrule
    \multicolumn{4}{l}{\textit{Category 2 --- Foundational 3D Designs}} \\
    D0  & CATCH Reference Design         & CPU + MEM (stack) + GPU on interposer
        & Baseline from original CATCH files \\
    D5  & Simple CPU + Memory Stack      & MEM stacked on CS
        & Basic 3D stacking \\
    D6  & HBM-Style Memory Stack         & 4$\times$MEM stacked on TSV
        & Deep vertical stack depth \\
    D7  & Double-Sided Stacking          & CS on face + MEM on back of TSV
        & Face/back connectivity \\
    D8  & 3D Partitioned SoC             & CL stacked on CL
        & 3D partitioning benefits \\
    \midrule
    \multicolumn{4}{l}{\textit{Category 3 --- Complex Hybrid Architectures}} \\
    D9  & Classic Accelerator            & SSi + CL + 2$\times$(4$\times$MEM/TSV)
        & Realistic GPU+HBM topology \\
    D10 & Multi-Stack System             & SO + [CS+MEM] + [ANA]
        & Multiple independent stacks \\
    D11 & Pyramid Architecture           & CL on SO, 2$\times$CS stacked on CL
        & Inverted pyramid stacking \\
    D12 & Buried Bridge Chip             & 2$\times$CL on SO + buried SSi bridge
        & EMIB-like interconnect \\
    D13 & Full System Integration        & Mixed multi-stack with all templates
        & Maximum heterogeneity \\
    \midrule
    \multicolumn{4}{l}{\textit{Category 4 --- Model Stress-Test Variants}} \\
    D14 & Routing Congestion Failure     & D3 variant, SSi with 1 routing layer
        & Congestion check \\
    D15 & NRE Cost Dominance             & D10 variant, CS at 10K qty vs.\ others at 10M
        & NRE amortization \\
    D16 & Assembly Process Comparison    & D2 run with individual vs.\ simultaneous bonding
        & Assembly cost model \\
    D17 & 2.5D Memory-Compute Grid       & 4$\times$MEM + 4$\times$CS on interposer
        & Square tiling arrangement \\
    D18 & 4$\times$ Analog on Interposer & 4$\times$ANA on SSi
        & Small-count analog stacking cost \\
    D19 & 64$\times$ Analog on Interposer& 64$\times$ANA on SSi
        & Large-count analog cost scaling \\
    \bottomrule
  \end{tabular}
\end{table*}
 
\textbf{Category 1} D1--D4 establish foundational 2.5D configurations, ranging from a simple heterogeneous pair to multi-chiplet, heterogeneous, and I/O-dominated systems.

\textbf{Category 2} D0, D5--D8 introduce vertical integration through simple stacking, deep TSV-based stacks, double-sided connectivity, and 3D partitioning.
 
\textbf{Category 3} D9--D13 combine these constructs into more complex heterogeneous systems, including HBM-style stacks, multiple independent stacks, non-uniform hierarchies, buried bridges, and full-system integration.
 
\textbf{Category 4} D14--D19 deliberately perturb selected design parameters to exercise congestion, NRE amortization, assembly-process selection, placement regularity, and chiplet-count scaling.

\subsection{Open-Source Benchmark Package}

The 3Dblox repository is organized to provide not only testcase descriptions but also the supporting artifacts required to reproduce and extend them. The complete set of benchmark designs is publicly available through the Dryad repository (DOI: 10.5061/dryad.76hdr7tb6).

%% file: tex/4_conclusion.tex
\section{Conclusion}
\label{sec:conclusion}

In this paper, we have presented the open-source benchmark suite explicitly designed for 2.5D/3D heterogeneous integration research in physical design. Our suite provides 20 designs spanning foundational 2.5D configurations, basic 3D stacking topologies, complex hybrid architectures, and targeted stress-test variants. The benchmark suite provides a common set of heterogeneous 2.5D/3D structures that can be extended as new integration technologies and design problems emerge. By releasing the designs in 3Dblox together with supporting open-source physical-design collateral, we hope to enable reproducible comparison across academic and commercial 3D design methodologies.